\documentclass[aps,prl,twocolumn,numerical,superscriptaddress,nofootinbib,showpacs,showkeys,longbibliography]{revtex4-1}

\usepackage{bm}
\usepackage{graphicx,color}
\usepackage{amsmath,amssymb,amsfonts}
\usepackage[colorlinks=true,linkcolor=blue,plainpages=false]{hyperref}
\usepackage[utf8]{inputenc}

\newcommand{\be}{\begin{eqnarray}}
\newcommand{\ee}{\end{eqnarray}}

\def\v2{\mbox{$v_2$}}

\def\eq{{\,=\,}}

\usepackage{ulem}
\usepackage{color}

\begin{document}


\title{System-size dependence of the viscous attenuation of anisotropic flow \\ in p+Pb and Pb+Pb collisions at LHC energies}

\author{Peifeng Liu} 
\affiliation{Department of Chemistry, 
Stony Brook University, \\
Stony Brook, NY, 11794-3400, USA}
\affiliation{Department of Physics and Astronomy, Stony Brook University, \\
Stony Brook, NY, 11794-3800}
\author{ Roy~A.~Lacey}
\email[E-mail: ]{Roy.Lacey@Stonybrook.edu}
\affiliation{Department of Chemistry, 
Stony Brook University, \\
Stony Brook, NY, 11794-3400, USA}
\affiliation{Department of Physics and Astronomy, Stony Brook University, \\
Stony Brook, NY, 11794-3800}
%
%
%
%



\date{\today}

\begin{abstract}
The elliptic and triangular flow coefficients ($\mathrm{v_n, \, n=2,3}$) measured 
in Pb+Pb ($\sqrt{s_{_{\rm NN}}} = 2.76$ TeV) and p+Pb ($\sqrt{s_{_{\rm NN}}} = 5.02$ TeV) collisions, 
are studied as a function of initial-state eccentricity ($\varepsilon_n$), and dimensionless size 
characterized by the cube root of the mid-rapidity charged hadron multiplicity 
density $\mathrm{\left< N_{ch} \right>^{1/3}}$. The results indicate that the influence 
of eccentricity ($\mathrm{v_n} \propto \varepsilon_n$) observed for large $\mathrm{\left< N_{ch} \right>}$,
is superseded by the effects of viscous attenuation for small $\mathrm{\left< N_{ch} \right>}$, irrespective
of the colliding species. Strikingly similar acoustic scaling patterns of 
exponential viscous modulation, with a damping rate proportional to $\mathrm{n^2}$ and inversely 
proportional to the dimensionless size, are observed for the eccentricity-scaled coefficients for the 
two sets of colliding species.  
The resulting scaling parameters suggest that, contrary to current predilections, 
the patterns of viscous attenuation, as well as the specific shear viscosity $\left<\frac{\eta}{s}(\text{T})\right>$ 
for the matter created in p+Pb and Pb+Pb collisions, are comparable.
\end{abstract}

\pacs{ }

\date{\today}

\maketitle


Relativistic heavy-ion collisions can lead to the production 
of high energy density domains of strongly interacting matter with an anisotropic 
transverse energy density profile. The ensuing expansion and hadronization
of this so-called fireball of dense partonic matter, results in the production of particles 
with an azimuthal anisotropy that reflects the viscous hydrodynamic response to the 
initial anisotropic energy density profile 
\cite{Teaney:2003kp,Romatschke:2007mq,Song:2010mg,Staig:2011wj,Schenke:2011tv,
Bozek:2011if,Gardim:2012yp,Shuryak:2013ke,Qian:2016fpi,McDonald:2016vlt,Bernhard:2016tnd}. 

The shape of this profile $\rho_e(r,\varphi)$, can be characterized 
by the complex eccentricity vectors \cite{Alver:2010dn,Petersen:2010cw,
Lacey:2010hw,Teaney:2010vd,Qiu:2011iv}:
\begin{eqnarray}
\mathrm{\mathcal{E}_n  \equiv \varepsilon_n e^{in\Phi_n} \equiv 
  - \frac{\int d^2r_\perp\, r^m\,e^{in\varphi}\, \rho_e(r,\varphi)}
           {\int d^2r_\perp\, r^m\,\rho_e(r,\varphi)}},                                                       
\label{epsdef1}
\end{eqnarray}
where $\mathrm{\varepsilon_n = {\left< \left| \mathcal{E}_n \right|^2 \right>}^{1/2}}$ and $\mathrm{\Phi_n}$ denote 
the magnitude and azimuthal direction of the $\mathrm{n^{th}}$-order eccentricity vector 
which fluctuates from event to event;
$\mathrm{m\eq{n}}$ for $\mathrm{n{\geq\,}2}$ and $\mathrm{m\eq3}$ for 
$\mathrm{n\eq1}$ \cite{Teaney:2010vd,Bhalerao:2014xra,Yan:2015jma}. 

The anisotropic flow which derives from $\mathrm{\varepsilon_n}$, results in an azimuthal 
asymmetry of the measured single-particle distribution.
Thus, it can be quantified by the complex flow vectors \cite{Ollitrault:1992bk,Luzum:2011mm,Teaney:2012ke}:
\begin{equation}
 \mathrm{V_n  \equiv v_ne^{in\Psi_n} \equiv \{e^{in\phi}\}}, \ \ \mathrm{v_n = {\left< \left| V_n \right|^2 \right>}^{1/2}},
\label{Vndef}
\end{equation}
where $\phi$ denotes the azimuthal angle around the beam direction, of a particle emitted 
in the collision, $\{\dots\}$ denotes the average over all particles emitted in the event, 
and $\mathrm{v_n}$ and $\mathrm{\Psi_n}$ denote 
the magnitude and azimuthal direction of the $\mathrm{n^{th}}$-order harmonic flow vector 
which also fluctuates from event to event. 

Viscous hydrodynamical model investigations show that $\mathrm{v_n \propto \varepsilon_n}$ for  
elliptic and triangular flow ($\mathrm{n=2 \: and \: 3}$) for the ``large'' and moderate-sized 
systems produced in central and mid-central heavy ion 
collisions \cite{Qiu:2011iv,Fu:2015wba,Niemi:2015qia,Noronha-Hostler:2015dbi,Bernhard:2016tnd}.
They also indicate that the temperature dependent specific shear 
viscosity (i.e., the ratio of shear viscosity to entropy density $\frac{\eta}{s}(\text{T})$)
of the partonic medium produced in the collisions, serve to attenuate the magnitude of 
$\mathrm{v_n}$ and consequently the ratio $\mathrm{v_n/\varepsilon_n}$. 
Thus, viscous hydrodynamical model comparisons to $\mathrm{v_n}$ measurements have been,
and continue to be an important avenue to estimate  
the value of $\frac{\eta}{s}(\text{T})$  \cite{Hirano:2005xf,Romatschke:2007mq,Song:2010mg,Schenke:2010rr,Bozek:2010wt,
Qiu:2011iv,Schenke:2011tv,Niemi:2012ry,Gardim:2012yp,McDonald:2016vlt,Bernhard:2016tnd} for the 
partonic matter produced in these large to moderate-sized systems. 

For the small systems produced in peripheral heavy ion collisions and light-heavy ion 
collisions (eg. proton–nucleus collisions), there has been a pervasive predilection 
that collective flow does not develop because viscous hydrodynamically-driven 
expansion breaks down. In part, this notion stems from the expectation that 
microscopic scales such as the mean free path, are probably similar to the geometric size 
of these systems. Thus, the presence of the large gradients inherent to small systems, could 
excite non-hydrodynamic modes or render invalid, the hydrodynamic 
gradient expansion  \cite{Florkowski:2016zsi,Denicol:2012cn} required 
to accurately characterize the viscous hydrodynamic response.
It has also been argued that alternative mechanistic scenarios, such as initial state 
correlations \cite{Dusling:2012iga,Kovchegov:2014yza,Kovner:2012jm}, could account 
for the azimuthal anisotropy observed for these small systems. A decisive validation of such scenarios 
would make the question of the validity of viscous hydrodynamical evolution in small systems 
a moot point.

Recent experimental measurements at both RHIC \cite{Adare:2013piz,Adare:2015ctn} and 
the LHC \cite{Chatrchyan:2013nka,Abelev:2012ola,Aad:2012gla,Aaboud:2017acw}, 
have given strong indications for collective anisotropic flow in the small systems
produced in peripheral heavy ion and light-heavy ion collisions. 
Several attempts have also been made to reconcile these measurements with 
viscous hydrodynamical evolution of the high energy-density strongly 
interacting matter which comprise these systems \cite{Bozek:2012gr,Bzdak:2013zma,Schenke:2014zha,Habich:2015rtj}.
However, the disparate influence of initial-state eccentricity, system size and $\frac{\eta}{s}(\text{T})$ has not 
been fully charted. Currently, it is also unclear as to what constitutes a good experimental measure of 
the size of these systems, as well as how small they can be, and still hydrodynamize. 
For the latter, numerical simulations in strongly interacting theories suggests that hydrodynamics remains 
applicable when the system size ($\mathrm{R}$) is of ${\cal O}(1/\mathrm{T})$ -- the inverse temperature.
That is, when the dimensionless size $\mathrm{RT} \sim 1$ \cite{Chesler:2016ceu}. 

The respective influence of $\mathrm{\varepsilon_n}$, system size and $\frac{\eta}{s}(\text{T})$ on $\mathrm{v_n}$, 
can be studied within an acoustic model framework, akin to that  of 
relativistic viscous hydrodynamics \cite{Lacey:2011ug,Lacey:2013is,Shuryak:2013ke,Lacey:2013eia,Liu:2018hjh}. 
Within this framework, the viscous attenuation of $\mathrm{v_n}$ for a given 
mean transverse momentum $\mathrm{\left< p_T \right>}$ and centrality selection $\mathrm{cent}$, can be expressed as  
\cite{Staig:2011wj,Lacey:2011ug,Lacey:2013is,Shuryak:2013ke,Lacey:2013eia,Liu:2018hjh}:
\begin{eqnarray}
\mathrm{\frac{v_n}{\varepsilon_n}} &\propto& \mathrm{\exp{\left(-n^2 \beta \frac{1}{{RT}} \right)} \quad n=2,3},
\label{a_damp-L}
\end{eqnarray}
where $\beta \propto \frac{\eta}{s}(\text{T}))$ and $\mathrm{R}$ characterizes the  
geometric size of the medium produced in the collision. For a given centrality selection, 
the dimensionless size $\mathrm{RT \propto \left< N_{ch}\right>^{1/3}}$, where $\mathrm{\left< N_{ch} \right>}$ 
is the mid-rapidity charged hadron multiplicity density \cite{Lacey:2016hqy}.
The magnitude of $\mathrm{\left< N_{ch}\right>}$ is observed to factorize into a contribution 
proportional to the number of quark participant pairs $\mathrm{N_{qpp}}$ and a contribution proportional 
to $\mathrm{\log{(\sqrt{s_{_{NN}}})}}$ \cite{Lacey:2016hqy};
\begin{eqnarray}
\mathrm{\left< N_{ch} \right> = N_{qpp}[b_{AA} + m_{AA}\log(\sqrt{s_{_{NN}}})]^3} \nonumber \\ 
                  \mathrm{ b_{AA} = 0.530 \pm 0.008  \quad m_{AA} = 0.258 \pm 0.004 }.
\label{mult}
\end{eqnarray}
Thus, for a given $\mathrm{\sqrt{s_{_{NN}}}}$, variations in the magnitude 
of $\mathrm{\left< N_{ch} \right>}$ largely reflects a change in $\mathrm{N_{qpp}}$.
Indeed, for the simplifying assumption that $\mathrm{R \propto (N_{qpp})^{1/3}}$, 
Fig.~5 of Ref.~\cite{Lacey:2016hqy} shows that the ratio $\mathrm{[\left< N_{ch} \right>/(N_{qpp})]^{1/3}}$
is independent of $\mathrm{R \propto (N_{qpp})^{1/3}}$. This gives a further indication 
that, for a fixed value of $\mathrm{\sqrt{s_{_{NN}}}}$, variations in the magnitude of 
$\mathrm{RT}$ are not accompanied by significant 
temperature changes. In contrast, a $\mathrm{\sqrt{s_{_{NN}}}}$-driven variation of 
$\mathrm{RT}$ for a fixed value of $\mathrm{N_{qpp}}$, would result in significant 
temperature changes.

Equation \ref{a_damp-L} suggests characteristic linear dependencies for $\mathrm{\ln(v_n/\varepsilon_n)}$  
and $\mathrm{\ln[(v_3/\varepsilon_3)/(v_2/\varepsilon_2)]}$ on $\mathrm{\left< N_{ch} \right>^{-1/3}}$, 
with slopes that reflect the quadratic viscous attenuation prefactors for 
$\beta$; these combined features are termed acoustic scaling. 
Since $\mathrm{\left< N_{ch} \right>}$ is small for peripheral heavy ion collisions 
and light-heavy ion collisions, Eq.~\ref{a_damp-L} also suggests that an uncharacteristically 
large viscous suppression of $\mathrm{v_n}$ (relative to that for large systems) is to 
be expected for the small systems of dimensionless size $\mathrm{RT \propto \left< N_{ch}\right>^{1/3}}$,
irrespective of the colliding species. 

An observed similarity in the slopes for 
$\mathrm{\ln[(v_3/\varepsilon_3)/(v_2/\varepsilon_2)]}$ vs. $\mathrm{\left< N_{ch} \right>^{-1/3}}$, for 
both small and large systems, would not only confirm that $\mathrm{\left< N_{ch}\right>^{1/3}}$
is a good measure for the dimensionless size, but also provide strong evidence that 
$\left< \frac{\eta}{s}(\text{T}) \right>$ for the medium produced in these systems are comparable.
Thus, the validation of acoustic scaling for $\mathrm{v_n/\varepsilon_n}$ across the full range 
of dimensionless sizes afforded in Pb+Pb and p+Pb collisions, could provide further constraints 
for the range of applicability of viscous hydrodynamics, as well as aid its utility 
for precision extraction of $\frac{\eta}{s}(\text{T})$. 
%
\begin{figure*}
\includegraphics[width=0.90\linewidth]{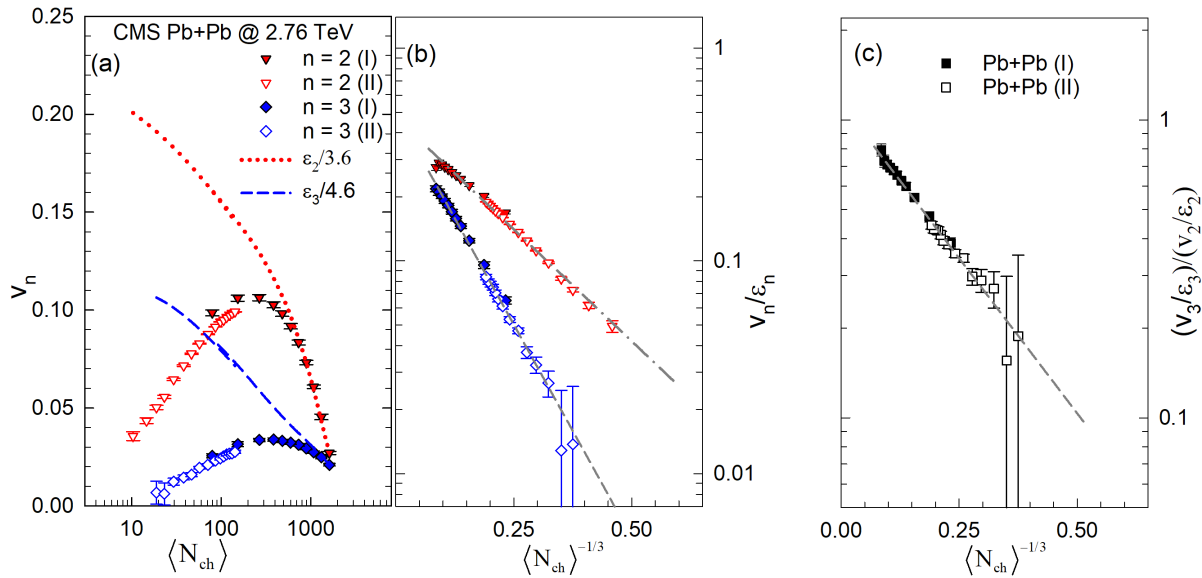}
\caption{(a) $\mathrm{v_n}$ and $\mathrm{\varepsilon_n}$ vs. $\mathrm{\left< N_{ch} \right>}$ 
for Pb+Pb collisions at $\sqrt{s_{_{\rm NN}}} = 2.76$. The filled and open symbols indicate 
centrality selected (I) and $\mathrm{N_{ch}}$-selected (II) measurements 
respectively. The dotted and dashed curves show $\mathrm{\varepsilon_n}$ values which are 
normalized to $\mathrm{v_n}$ at $\mathrm{\left< N_{ch} \right> \sim 1600}$; 
(b) $\mathrm{v_n/\varepsilon_n}$ vs. $\mathrm{\left< N_{ch} \right>^{-1/3}}$ for the data shown in panel (a).
The dashed lines represent fits to the eccentricity-scaled data following Eq.~\ref{a_damp-L};
(c) $\mathrm{[(v_3/\varepsilon_3)/(v_2/\varepsilon_2)]}$ vs. $\mathrm{\left< N_{ch} \right>^{-1/3}}$
for the data shown in panel (b). The dashed line represents a fit to the data following Eq.~\ref{a_damp-L}.  
}
\label{fig:1}
\end{figure*} 
%

In this letter, we use recent $\mathrm{p_T}$-integrated ($\mathrm{0.3 < p_T <3}$~GeV/c) 
$\mathrm{v_2}$ and $\mathrm{v_3}$ measurements in Pb+Pb ($\sqrt{s_{NN}}$ = 2.76 TeV) and 
p+Pb ($\sqrt{s_{NN}}$ = 5.02 TeV) collisions, to explore validation tests for  
acoustic scaling of $\mathrm{v_n/\varepsilon_n}$ and the ratio $\mathrm{(v_3/\varepsilon_3)/(v_2/\varepsilon_2)}$. 
We find that these tests provide crucial insight on the influence of system size on 
viscous hydrodynamic-like evolution.
Additionally, they allow a direct comparison of $\left<\frac{\eta}{s}(\text{T})\right>$ for the hot and dense 
medium produced in p+Pb and Pb+Pb collisions, over the full range of system sizes characterized by the 
dimensionless size $\mathrm{RT \propto \left< N_{ch}\right>^{1/3}}$. 

The data employed in this work are taken from the CMS centrality selected flow measurements 
for Pb+Pb collisions at $\sqrt{s_{NN}}$ = 2.76 TeV with the event plane 
method \cite{Chatrchyan:2012ta,Chatrchyan:2013kba}
and the CMS $\mathrm{N_{ch}}$-selected flow measurements for Pb+Pb collisions 
at $\sqrt{s_{NN}}$ = 2.76 TeV and p+Pb collisions 
at $\sqrt{s_{NN}}$ = 5.02 TeV \cite{Chatrchyan:2013nka}. 
The $\mathrm{N_{ch}}$ selections for the latter measurements were made for 2.4 units 
of pseudorapidity (i.e., $\mathrm{|\eta| < 2.4}$). Consequently, the efficiency corrected 
values were scaled to obtain $\mathrm{\left< N_{ch} \right>}$ for one unit of $\eta$, to ensure 
consistency with the centrality dependent measurements. The requisite $\mathrm{\left< N_{ch} \right>}$ values 
for the centrality selected measurements were obtained from the CMS multiplicity density 
measurements \cite{Chatrchyan:2011pb}.

The necessary $\mathrm{\left< N_{ch} \right>}$ dependent eccentricities were calculated 
following the procedure outlined in Eq.~\ref{epsdef1}, with the aid of a Monte Carlo 
quark-Glauber model (MC-qGlauber) with fluctuating initial conditions \cite{Pifeng-Liu}.
The model, which is based on the commonly used MC-Glauber model \cite{Miller:2007ri,*Alver:2006wh}, 
was used to compute the number of quark participants $\mathrm{Nq_{part}(cent)}$, 
$\mathrm{\varepsilon_n(cent)}$ and $\mathrm{\varepsilon_n(\left< N_{ch} \right>)}$ 
from the two-dimensional profile of the density of sources in the transverse  
plane $\rho_s(\mathbf{r_{\perp}})$ \cite{Pifeng-Liu,Lacey:2010hw,Teaney:2012ke}. 
The model takes account of the finite size of the nucleon, the wounding profile of the nucleon,
the distribution of quarks inside the nucleon and quark cross sections which reproduce 
the NN inelastic cross section at $\sqrt{s_{NN}}$ = 2.76 and 5.02 TeV. A systematic uncertainty 
of 2-5\% was estimated for the eccentricities from variations of the model parameters.

Validation tests for acoustic scaling were performed 
by plotting $\mathrm{v_n/\varepsilon_n}$ and $\mathrm{(v_3/\varepsilon_3)/(v_2/\varepsilon_2)}$ 
vs. $\mathrm{\left< N_{ch} \right>^{-1/3}}$ respectively, followed by evaluations for the expected patterns 
of exponential viscous attenuation, and the relative viscous attenuation $\beta$-prefactors 
indicated in Eq.~\ref{a_damp-L}.

Figure \ref{fig:1}(a) shows the $\mathrm{\left< N_{ch} \right>}$ dependence of 
$\mathrm{v_2}$ and $\mathrm{v_3}$ for the combined data sets for Pb+Pb collisions,
as well as the $\mathrm{\left< N_{ch} \right>}$ dependence of 
$\mathrm{\varepsilon_2}$ and $\mathrm{\varepsilon_3}$. To facilitate a comparison between $\mathrm{v_n}$
and $\mathrm{\varepsilon_n}$, the $\mathrm{\varepsilon_n}$ values are divided by a factor 
(indicated in the figure) to normalize $\mathrm{v_n}$ and $\mathrm{\varepsilon_n}$ at 
$\mathrm{\left< N_{ch} \right> \sim 1600}$.
Fig.~\ref{fig:1}(a) shows that, for $\mathrm{\left< N_{ch} \right> \agt 400}$, 
$\mathrm{v_n}$ follows the trend for $\mathrm{\varepsilon_n}$, i.e., $\mathrm{v_n} \propto \mathrm{\varepsilon_n}$. 
However, for $\mathrm{\left< N_{ch} \right>} \alt 400$,
the $\mathrm{v_n}$ trend is opposite to that for $\mathrm{\varepsilon_n}$. We attribute this 
unmistakable difference in trend, to the very large viscous attenuation effects which result for 
small system sizes, i.e., small dimensionless sizes. That is, for small values of $\mathrm{RT}$, 
the expectation that $\mathrm{v_n \propto \varepsilon_n}$ is supplanted by the dominating effects of 
the exponential viscous attenuation indicated in Eq.~\ref{a_damp-L}. 

This pattern of exponential viscous attenuation is made transparent in Figs.~\ref{fig:1}(b) and (c),
which show the telltale acoustic scaling patterns of a characteristic linear dependence 
of $\mathrm{\ln(v_n/\varepsilon_n)}$ and $\mathrm{\ln[(v_3/\varepsilon_3)/(v_2/\varepsilon_2)]}$ 
on $\mathrm{(N_{ch})^{-1/3}}$ (respectively), with slope factors which reflect the 
$\mathrm{n^2}$ dependence on harmonic number. Note that the $\beta$-prefactors 
(indicated in Eq.~\ref{a_damp-L}) for $\mathrm{\ln(v_n/\varepsilon_n)}$ vs. $\mathrm{(N_{ch})^{-1/3}}$
are 4 and 9 for n = 2 and 3 respectively, and the prefactor for 
$\mathrm{\ln[(v_3/\varepsilon_3)/(v_2/\varepsilon_2)]}$ vs. $\mathrm{(N_{ch})^{-1/3}}$ is 5.
Note as well that the ratio $\mathrm{(v_3/\varepsilon_3)/(v_2/\varepsilon_2)}$ leads to 
a significant reduction in the influence of possible $\mathrm{p_T}$-dependent viscous 
effects \cite{Liu:2018hjh} for the $\mathrm{\left< p_T \right>}$ values of interest.
%
\begin{figure*}[t]
\includegraphics[width=0.90\linewidth]{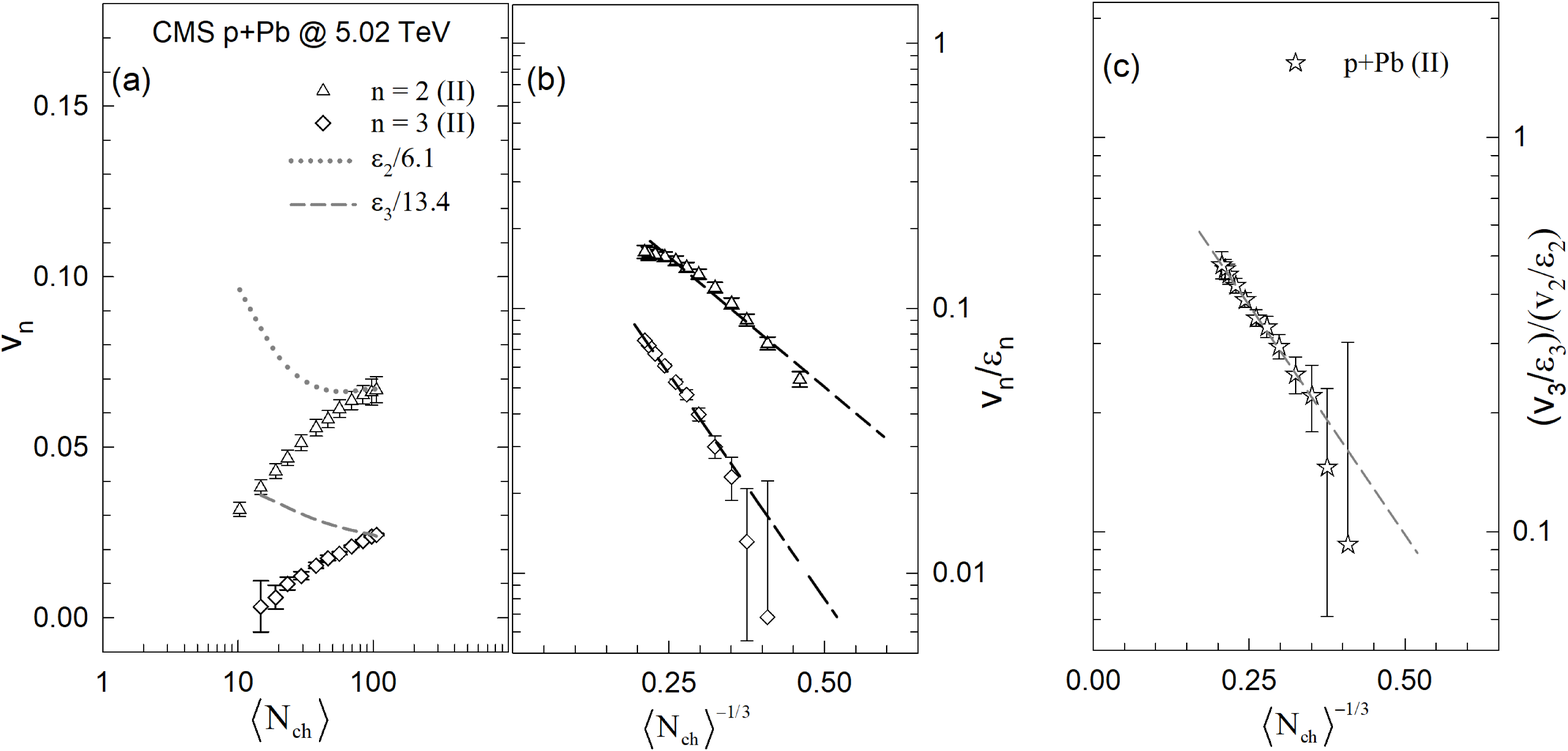}
\caption{(a) $\mathrm{v_n}$ and $\mathrm{\varepsilon_n}$ vs. $\mathrm{\left< N_{ch} \right>}$ 
for $\mathrm{N_{ch}}$-selected (II) measurements of $\mathrm{v_n}$ in 
p+Pb collisions at $\sqrt{s_{_{\rm NN}}} = 5.02$~TeV. The dotted and dashed curves 
show $\mathrm{\varepsilon_n}$ values which are normalized to $\mathrm{v_n}$ 
at $\mathrm{\left< N_{ch} \right> \sim 60}$; 
(b) $\mathrm{v_n/\varepsilon_n}$ vs. $\mathrm{\left< N_{ch} \right>^{-1/3}}$ for the data shown in panel (a).
The dashed lines represent fits to the eccentricity-scaled data following Eq.~\ref{a_damp-L};
(c) $\mathrm{\ln[(v_3/\varepsilon_3)/(v_2/\varepsilon_2)]}$ vs. $\mathrm{\left< N_{ch} \right>^{-1/3}}$
for the data shown in panel (b). The dashed line represents a fit to the data 
following Eq.~\ref{a_damp-L}
}
\label{fig:2}
\end{figure*}
%

The dashed lines in Figs.~\ref{fig:1}(b) and (c) represent fits to the eccentricity-scaled 
data following Eq.~\ref{a_damp-L}. They indicate relatively good fits which suggest a size-independent 
$\left<\frac{\eta}{s}(\text{T})\right>$ value for the full range of system sizes spanned by the 
combined data sets for these Pb+Pb collisions. This observation is consistent with the earlier expectation 
that variations in the magnitude of $\mathrm{RT}$ at a fixed value of $\mathrm{\sqrt{s_{_{NN}}}}$,
are not accompanied by significant temperature variations. Moreover, a weak temperature dependence 
of $\frac{\eta}{s}(\text{T})$ has been indicated in recent viscous hydrodynamical 
calculations \cite{Bernhard:2016tnd}.
This implied size-independence of $\left<\frac{\eta}{s}(\text{T})\right>$, does not 
preclude viscous attenuation which is significantly larger in small systems than in large 
systems (cf. Figs.~\ref{fig:1}(b) and (c)).

The results obtained for the $\mathrm{N_{ch}}$-selected measurements for 
p+Pb are shown in Fig. \ref{fig:2}. They indicate strikingly similar patterns to the ones shown in
Fig.~\ref{fig:1} for the $\mathrm{N_{ch}}$-selected Pb+Pb data, albeit with different 
magnitudes for $\mathrm{v_n}$ and $\mathrm{\varepsilon_n}$.
Fig.~\ref{fig:2}(a) shows that the $\mathrm{\left< N_{ch} \right>}$ dependence of $\mathrm{v_n}$ 
is opposite to the trend for $\mathrm{\varepsilon_n}$ over the full range of the measurements.
Here, the $\mathrm{\varepsilon_n}$ values are also divided by the indicated factors, to 
normalize $\mathrm{v_n}$ and $\mathrm{\varepsilon_n}$ at $\mathrm{\left< N_{ch} \right> \sim 60}$.
Note that the $\mathrm{\left< N_{ch} \right>}$ values in Fig.~\ref{fig:2} are for one 
unit of pseudorapidity. The statistical significance of the $\mathrm{\varepsilon_n}$ values, precluded
a comparison to a few data points at larger $\mathrm{\left< N_{ch} \right>}$ corresponding
to the top 1\% fraction of the events.  
The observed trends confirm that the very large viscous attenuation effects, 
previously identified for the small systems created in Pb+Pb collisions (c.f. Fig.~\ref{fig:1}), 
are also present in these p+Pb collisions, for comparable $\mathrm{\left< N_{ch} \right>}$. 
%
\begin{figure}[h]
\includegraphics[width=0.70\linewidth]{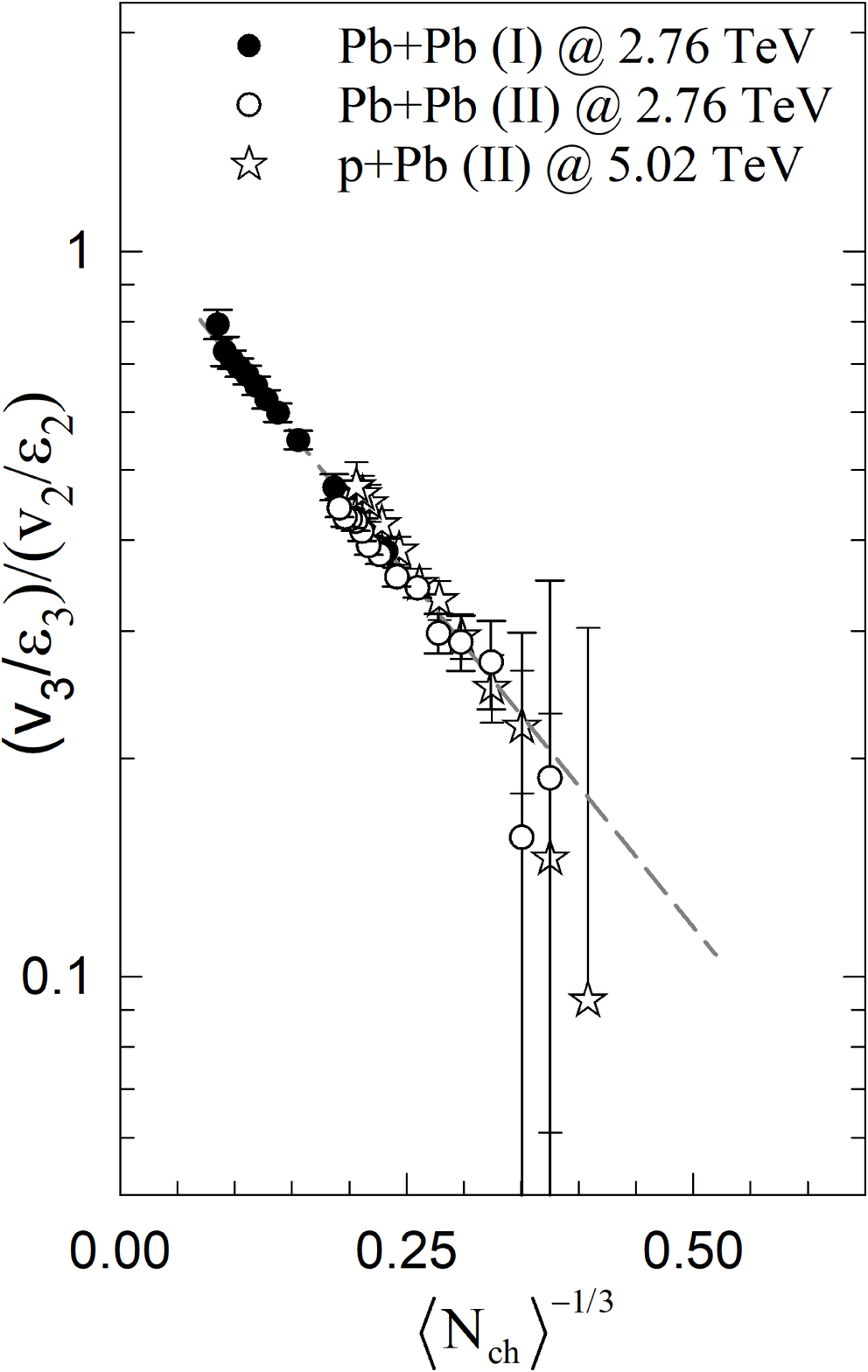}
%
\caption{ Comparison of $\mathrm{\ln[(v_3/\varepsilon_3)/(v_2/\varepsilon_2)]}$ vs. $\mathrm{\left< N_{ch} \right>^{-1/3}}$
for centrality selected (I) and $\mathrm{N_{ch}}$-selected (II) Pb+Pb measurements, 
and $\mathrm{N_{ch}}$-selected (II) p+Pb measurements. The dashed line represents a fit 
to the combined data sets, following Eq.~\ref{a_damp-L}.
}
\label{fig:3}
\end{figure} 
%

The revealing acoustic scaling patterns of a characteristic linear dependence 
of $\mathrm{\ln(v_n/\varepsilon_n)}$ and $\mathrm{\ln[(v_3/\varepsilon_3)/(v_2/\varepsilon_2)]}$ 
on $\mathrm{(N_{ch})^{-1/3}}$ (respectively), with slope factors which reflect the 
$\mathrm{n^2}$ dependence on harmonic number, are also apparent in Figs.~\ref{fig:2}(b) and (c).
They are strikingly similar to the scaling patterns previously observed in Figs.~\ref{fig:1}(b) and (c)
for Pb+Pb collisions. The indicated dashed lines, which represent fits to the eccentricity-scaled 
data (following Eq.~\ref{a_damp-L}), also suggest a common $\left<\frac{\eta}{s}(\text{T})\right>$ over 
the full range of dimensionless sizes characterized by $\mathrm{\left< N_{ch} \right>}$ in these p+Pb collisions. 
Here, it should be emphasized again that this implied size-independence 
of $\left<\frac{\eta}{s}(\text{T})\right>$ for the medium produced in p+Pb systems of varying sizes, 
are fully compatible with the large size-dependent viscous attenuation effects produced in these collisions.

The size dependence of the ratio of the eccentricity-scaled coefficients 
$\mathrm{(v_3/\varepsilon_3)/(v_2/\varepsilon_2)}$, for the 
p+Pb and Pb+Pb measurements are compared in Fig.~\ref{fig:3}. 
Note that this ratio leads to a significant reduction in the influence of 
possible $\mathrm{p_T}$-dependent viscous effects \cite{Liu:2018hjh} which 
could be different for p+Pb and Pb+Pb collisions.
The comparison indicates strikingly similar magnitudes and slope trends for both 
sets of measurements. Note that it is the slope which carries information about 
$\left<\frac{\eta}{s}(\text{T})\right>$, not the magnitude. The latter is not required
to be similar for the two sets of colliding species to have comparable values 
of $\left<\frac{\eta}{s}(\text{T})\right>$. The dashed line, which represent the results 
from a fit to the combined data sets, indicate that, within an uncertainty of 
$\sim 6$\%, a single slope value $\beta = 0.83 \pm 0.05$, can account for the wealth of the 
combined measurements. To estimate the overall fit uncertainty, independent fits were performed 
for each data set. 

The observed similarity between the p+Pb and Pb+Pb $\beta$ values, 
suggests that $\left< \frac{\eta}{s}(\text{T}) \right>$ for the medium created in p+Pb and Pb+Pb 
collisions are comparable. However, a further calibration that maps 
the extracted value of $\beta$ on to a quantifiable value for $\left< \frac{\eta}{s}(\text{T})\right>$
would be required. An appropriately constrained set of viscous hydrodynamical calculations, 
tuned to reproduce the results shown in Figs. \ref{fig:1} - \ref{fig:3}, could provide such a calibration 
to give a relatively precise estimate, as well as simultaneous validation 
of the initial-state eccentricity spectrum for these collisions.


In summary, we have presented a detailed phenomenological investigation of the 
influence of dimensionless size $\mathrm{RT \propto \left< N_{ch} \right>^{1/3}}$, 
on the viscous attenuation of the elliptic and triangular 
flow coefficients measured in Pb+Pb ($\sqrt{s_{_{\rm NN}}} = 2.76$ TeV) and 
p+Pb ($\sqrt{s_{_{\rm NN}}} = 5.02$ TeV) collisions. We find that, for 
for small $\mathrm{\left< N_{ch} \right>}$ (small dimensionless size),
the magnitude of the flow coefficients are dominated by the effects of size-driven 
viscous attenuation in both p+Pb and Pb+Pb collisions. 
Strikingly similar acoustic scaling patterns of exponential viscous modulation, with a damping rate 
proportional to $\mathrm{n^2}$ and inversely proportional to the dimensionless size, are 
observed for both the p+Pb and Pb+Pb eccentricity-scaled coefficients.
Such patterns suggest that the very large viscous attenuation effects,
apparent in the small systems created in p+Pb collisions are also present in Pb+Pb collisions 
of comparable $\mathrm{\left< N_{ch} \right>}$.
The scaling parameters for the ratio of the eccentricity-scaled $\mathrm{v_n}$ coefficients, 
further suggests comparable size-independent specific shear viscosities $\left<\frac{\eta}{s}(\text{T})\right>$ 
for the hot and dense matter produced in p+Pb and Pb+Pb collisions, contrary to current predilections.
These results could provide additional stringent constraints for the range of applicability of 
viscous hydrodynamics, as well as to aid its utility for precision extraction of 
the transport coefficients for hot and dense partonic matter.

{\bf Acknowledgments}
This research is supported by the US DOE under contract DE-FG02-87ER40331.A008. 
 




%
\bibliography{acoustic_pPb_PbPb-vn} 
\end{document}